\documentclass[nofootinbib,twocolumn,aps,prd,superscriptaddress,longbibliography]{revtex4-2}

\usepackage{amsmath,amssymb,bm}
\usepackage{graphicx}
\usepackage{hyperref}
\usepackage{xcolor}

\begin{document}

\title{The astrometric precision of incoherent pulsar timing arrays}

\author{Anna C. Tsai}
\email{anna.tsai@mail.utoronto.ca}
\affiliation{Canadian Institute for Theoretical Astrophysics, 60 St. George Street, Toronto, ON M5S 3H8, Canada}
\affiliation{Department of Physics, University of Toronto, 60 St. George Street, Toronto, ON M5S 1A7, Canada}
\affiliation{Dunlap Institute for Astronomy \& Astrophysics, University of Toronto, AB 120-50 St. George Street, Toronto, ON M5S 3H4, Canada}

\author{Dylan L. Jow}
\email{dylanjow@stanford.edu}
\affiliation{Kavli Institute for Particle Astrophysics and Cosmology, Stanford University, 452 Lomita Mall, Stanford, CA 94305-4085, USA}

\author{Ue-Li Pen}
\affiliation{Institute of Astronomy and Astrophysics, Academia Sinica, Astronomy-Mathematics Building, No. 1, Section 4, Roosevelt Road, Taipei 10617, Taiwan}
\affiliation{Canadian Institute for Theoretical Astrophysics, 60 St. George Street, Toronto, ON M5S 3H8, Canada}
\affiliation{Department of Physics, University of Toronto, 60 St. George Street, Toronto, ON M5S 1A7, Canada}
\affiliation{Dunlap Institute for Astronomy \& Astrophysics, University of Toronto, AB 120-50 St. George Street, Toronto, ON M5S 3H4, Canada}
\affiliation{Perimeter Institute for Theoretical Physics, 31 Caroline St. North, Waterloo, ON N2L 2Y5, Canada}
\affiliation{Canadian Institute for Advanced Research, CIFAR Program in Gravitation and Cosmology, MaRS Centre, West Tower, 661 University Ave., Suite 505, Toronto, ON M5G 1M1, Canada}

\begin{abstract}
We address the astrometric precision with which a pulsar timing array (PTA) can localize a single astrophysical gravitational-wave source, in the regime where pulsar distances are poorly constrained compared to the gravitational wavelength. Using simple matched-filter, map-making simulations, we calculate the astrometric precision of the intrinsic PTA residual response as a function of the signal-to-noise ratio (SNR), where the SNR is that of the source in the matched-filter map, and find that the median angular error decreases as $\sqrt{\ln 2/3}\,/[\mathrm{SNR}\sqrt{\ln\mathrm{SNR}}]$ over the relevant SNR range, with a pre-factor we derive from first principles. At an SNR of 6, this log term coming from the singular core of the convolution kernel produces an angular error approximately $1.4$ times smaller than for a Gaussian beam. We compare this sky-averaged maximum precision with the precision achieved across the sky by the pulsars of the International Pulsar Timing Array (IPTA) DR2. For a map with SNR of 6, the intrinsic array precision is $3.3^\circ$. For the IPTA configuration at the same SNR, the median precision over the sky is worse due to sparse sampling of the underlying pulsar residual response, approximately $8.9^\circ$; however, for sources in $14\%$ of the sky, the high local pulsar density yields a precision better than $3.3^\circ$.
\end{abstract}

\maketitle

\section{Introduction}

Pulsar timing arrays (PTAs) have now found evidence for a nanohertz gravitational-wave (GW) background, characterized by the Hellings--Downs angular correlation in the timing residuals of millisecond pulsars \cite{NANOGravdetection, EPTAdetection, Parkesdetection, CPTAdetection}. The leading interpretation is that this background is the superposition of GWs from a population of supermassive black hole binaries (SMBHBs). The immediate next goal is to disentangle the brightest of these binaries from the background.
Detecting such a source above the background would be valuable on its own. Localizing its position on the sky would be even more valuable, as precise localization enables host-galaxy identification, electromagnetic follow-up, and multimessenger studies of SMBHBs.

In this paper, we focus on \emph{astrometric precision}: how well a single source can be localized on the sky. Here, we take the term ``astrometric precision"  as distinct from ``angular resolution", which is the minimum separation at which two point sources can be distinguished. Our approach treats PTAs as an imaging GW telescope whose aperture is set by the distribution of pulsars on the sky \cite{Boyle2012}.
In this picture, astrometric precision depends on the pulsar configuration, the signal-to-noise ratio (SNR), and how much of the signal is used in the analysis. We must also consider the question of detector coherence. Each pulsar's timing residual contains an Earth term, which is common to all pulsars, and a pulsar term, whose phase depends on the distance to the pulsar. When pulsar distances are known to a precision better than the GW wavelength (roughly a parsec at nanohertz frequencies), the pulsar terms can be modeled coherently, which greatly improves localization \cite{Tsai2026}. For most pulsars in current arrays, however, the distance uncertainties are far larger than the GW wavelength, so the pulsar term cannot be used to improve localization significantly. In this work, we refer to an analysis that uses only the Earth term, with no pulsar distance information, as \emph{incoherent}. This approximates the regime in which current PTA analyses operate.

This paper is the third in a series on the imaging capabilities of PTAs. In Ref.~\cite{Jow2026} we develop a matched-filter map-making technique for incoherent PTAs and use it to bound the number of astrophysical GW sources that can be resolved at a map SNR of order unity --- a bound that is reached with a uniform distribution of roughly 20 or more pulsars. We find the angular resolution of an incoherent PTA to be $28.9^{\circ}$, set by the characteristic half width of the point spread function (PSF). In Ref.~\cite{Tsai2026}, we demonstrate a scaling relation for the angular localization of a single source in a coherent (i.e. distance-informed) analysis as a function of the number of pulsars with constrained distances. Here, we address the question of astrometric precision in the incoherent regime by asking how precisely a single source can be localized. We use the polarized matched-filter maps constructed in detail in Ref.~\cite{Jow2026} and simulate hundreds of noise realizations at a map SNR of 6. This roughly corresponds to a detection significance comparable to the current sensitivity of PTAs to a single source. We first recover the intrinsic astrometric precision of a PTA by simulating a fully dense, uniform pulsar array. We find the intrinsic localization at an SNR of 6 to be $3.3^\circ$. We then compare this limit with the precision achievable across the sky for the pulsars in IPTA DR2 and find that the median astrometric precision across the sky is $8.9^\circ$; however, sources can be localized to better than $3.3^{\circ}$ in the $14\%$ of the sky with highest pulsar density.

The paper is organized as follows. In Sec.~\ref{sec:maps} we review the matched-filter map-making technique and define our measure of astrometric precision. In Sec.~\ref{sec:uniform} we derive the intrinsic precision of an incoherent PTA and its scaling with SNR, and confirm it through simulations of dense, uniform pulsar arrays. In Sec.~\ref{sec:IPTA} we compare the intrinsic PTA precision with the precision across the sky of the IPTA DR2 configuration.

\section{Matched-filter maps}
\label{sec:maps}

We briefly review the construction of matched-filter (``dirty'') maps from
pulsar timing residuals; a full presentation is given in Ref.~\cite{Jow2026}.
We consider a monochromatic, circularly polarized gravitational wave of
frequency $f_0$ from a source in direction $\hat{n}$. We assume the pulsar
distances are poorly known, so the oscillatory pulsar term acts as an
additional source of noise and only the Earth term contributes coherently across
the array. The Fourier
amplitude of the timing residual of a pulsar at $\hat{r}$ is then
$\tilde\tau = -h\,\zeta^{L,R}(\hat{r};\hat{n})/(4\pi f_0)$, where the
dimensionless angular response is (up to an overall phase $e^{2\pi i f_0 t_0}$)
\begin{equation}
    \zeta^{L,R}(\hat{r};\hat{n}) = -\left(1+\hat{n}\cdot\hat{r}\right)e^{\mp 2i\psi},
    \label{eq:zeta}
\end{equation}
with $\psi = \arctan2\!\left(\hat{r}\cdot\hat{\phi}_{n},\,\hat{r}\cdot\hat{\theta}_{n}\right)$ as
the position angle of the pulsar about the source, measured from the
spherical basis vector $\hat\theta_n$ at $\hat n$. The amplitude of the
response varies slowly across the sky. Its phase, $2\psi$, comes from the
spin-2 polarization of the wave and winds twice around the source direction. Equation~\eqref{eq:zeta} samples an underlying, intrinsic PTA response map at each pulsar location, $\hat{r}$. This intrinsic response exhibits a singular core at the GW location (see Ref.~\cite{Jow2026} Fig.~5).

For a PTA of $N$ pulsars at positions $\hat{r}_i$, we pixelize the sky into
$M$ HEALPix pixels $\hat{n}_j$ \cite{HEALPix} (with resolution parameter $N_{\rm side}$ such that $M = 12 N_{\rm side} ^2$)
and construct the $N\times M$ aperture matrix
\begin{equation}
    A^{L}_{ij} = \zeta^{L}(\hat{r}_i;\hat{n}_j),
\end{equation}
whose $j$th column is the response of the array to a left circularly
polarized source at $\hat{n}_j$ ($A^R$ is defined analogously). Given the
vector of observed residuals $\mathbf{d}$ and the noise covariance $\Sigma$,
the left-handed dirty map is
\begin{equation}
    \mathbf{L} = (A^{L})^{\dagger}\,\Sigma^{-1}\,\mathbf{d}.
    \label{eq:dirty}
\end{equation}
Each element $L_j$ is the noise-weighted overlap between the data and the
response expected from a left-handed source at $\hat n_j$, i.e.,\ the output
of a matched filter for a point source at that pixel. In this work, we take
the pulsars to have identical white noise, $\Sigma = \sigma^2 \mathbb{I}$, so
that $\mathbf{L}\propto (A^L)^\dagger\mathbf{d}$. If the intrinsic PTA response were fully sampled (sampled by a fully dense, uniform array of pulsars), this map-making procedure would be equivalent to convolving the residual data with the intrinsic quadrupolar pulsar response. It is the intrinsic PTA response map, used as the convolution kernel, which yields improved astrometric precision over a Gaussian kernel.

We model the data as $\mathbf{d} = \mathbf{s} + \mathbf{n}$, where $\mathbf{s}$
is the response to a single source and $\mathbf{n}$ is complex
Gaussian noise in the residual space. The map is linear in the data, so the
dirty map is the sum of the noiseless point-spread function and a noise map
$(A^{L})^{\dagger}\mathbf{n}$. The noise map is correlated between pixels,
with covariance $(A^{L})^{\dagger}A^{L} \sigma^2$. We estimate the source position as the pixel of maximum $|L_j|$:
\begin{align}
\begin{split}
    &\hat{n}_{\rm max} = \hat{n}_{j^\ast}, \\
    &j^\ast = \arg\max_j |L_j|.
\end{split}
\end{align}
The astrometric error of a realization is the angular separation
\begin{equation}
    \gamma = \arccos(\hat{n}_{\rm max}\cdot\hat{n}),
\end{equation}
and we characterize the
precision by the median of $\gamma$ over noise realizations, which we denote $\gamma_{\rm med}$. We
define the SNR of a map as the peak of the noiseless map
divided by the RMS noise amplitude at the source pixel,
\begin{equation}
    \mathrm{SNR} = \frac{|L_{j^*}^{(s)}|}
    {\left\langle |(A^{L})^{\dagger}\mathbf{n}|^{2}_{j_{\rm GW}}\right\rangle^{1/2}},
    \label{eq:snr}
\end{equation}
where $L^{(s)} = (A^L)^\dagger\mathbf{s}$, $j_{\rm GW}$ is the pixel closest
to $\hat n$ , and the average is over noise realizations.

\section{Intrinsic astrometric precision of a PTA}
\label{sec:uniform}

In this section, we estimate the astrometric precision of the intrinsic PTA response by considering a dense, uniform pulsar array. This configuration gives the same precision for sources anywhere on the sky and represents the best achievable sky-averaged localization of an incoherent PTA. We focus on the intrinsic PTA response because, without prior knowledge of the source position and with no freedom to choose where pulsars lie, the sky-averaged precision is the relevant figure of merit for a blind search (see Sec.~\ref{sec:discussion}).

\subsection{Scaling relation}
\label{sec:scaling}

For a Gaussian PSF, the localization error scales with signal as $1/\mathrm{SNR}$ --- a factor of ten improvement in the signal to noise improves the astrometric precision by a factor of ten. Here we show that the singular core of the incoherent PTA response (Fig. 5 of Ref.~\cite{Jow2026}) improves this to $1/[\mathrm{SNR}\sqrt{\ln\mathrm{SNR}}]$ --- a factor of ten improvement in the signal to noise improves the astrometric precision by a factor of $\approx 15$. We first give a heuristic argument that shows how this additional logarithmic factor arises. Then, we derive the pre-factor with a Fisher-information calculation.

We begin with a heuristic argument for the slope of the scaling relation. For a single source placed at a pole for simplicity and detected with a given SNR in the matched-filter map, we divide the residual response map into $\mathrm{SNR}^2$ concentric, equal-area rings centered on the GW source position, where the first and last rings are spherical caps. For a dense, uniform array, each ring then contains the same number of pulsars and has $\mathrm{SNR}_k = 1$. In reality, the ring SNR varies with radius, being lower near the antipode; however, this effect only changes the constant and not the scaling. The outer radius $\theta_k$ of the $k$th ring is defined recursively by
\begin{equation}
    A_k = 2\pi\left(\cos\theta_{k-1} - \cos\theta_k\right) = \frac{4\pi}{\mathrm{SNR}^2},
\end{equation}
with $\theta_0 = 0$. Summing the areas of the first $k$ rings, the sum is a telescoping series, and evaluates to
\begin{equation}
    \sum_{j=1}^{k} A_j = 2\pi\left(1 - \cos\theta_k\right) = \frac{4\pi k}{\mathrm{SNR}^2},
\end{equation}
which gives $\theta_k = 2\arcsin\left(\sqrt{k}/\mathrm{SNR}\right) \approx 2\sqrt{k}/\mathrm{SNR}$ near the source.

For a single ring of pulsars, Ref.~\cite{Taylor2026} derives $\gamma_k \propto \theta_k/\mathrm{SNR}_k$, where $\gamma_k$ is the astrometric precision obtained from the matched-filter map of the $k$th ring alone. Assuming the noise in different rings is independent, the inverse variances add:
\begin{equation}
    \frac{1}{\gamma^2} \propto \sum_{k=1}^{\mathrm{SNR}^2}\frac{\mathrm{SNR}_k^2}{\theta_k^2}
    \propto \mathrm{SNR}^2 \sum_{k=1}^{\mathrm{SNR}^2}\frac{1}{k}
    \approx \mathrm{SNR}^2 \ln \mathrm{SNR},
\end{equation}
where in the last step we have used the asymptotic expansion of the harmonic number, $H_{\mathrm{SNR}^2} \approx 2\ln\mathrm{SNR} + \gamma_E$, where $\gamma_E$ is the Euler--Mascheroni constant. This gives the scaling relation
\begin{equation}
    \gamma \propto \frac{1}{\mathrm{SNR}\sqrt{\ln\mathrm{SNR}}}.
\end{equation}

The prefactor can be derived from first principles. Shifting the source by a small angle $\gamma$ changes the position angle of a pulsar at angular distance $\theta$ from the source by $\gamma\sin\psi\cot\theta$, and so rotates the spin-2 phase of its response, Eq.~\eqref{eq:zeta}. The angle $\theta$ itself also changes by $-\gamma\cos\psi$ --- using the spherical law of cosines --- which alters the amplitude by $1+\cos\theta$. For complex white noise of total variance $\sigma^2$, the per-axis Fisher information is $F = (2h^2/\sigma^2)\sum_i|\partial\zeta_i/\partial\gamma|^2$. The factor of 2 arises because each noise quadrature has variance $\sigma^2/2$. Replacing the sum by an integral over a uniform array, and normalizing with the SNR definition of Eq.~\eqref{eq:snr}, for which $\sum_i|\zeta_i|^2 = 4N/3$, gives
\begin{equation}
\begin{split}
    F &= \frac{3}{8}\,\mathrm{SNR}^2\int_\epsilon^\pi \sin\theta\,\bigl[\sin^2\theta + 4(1+\cos\theta)^2\cot^2\theta\bigr]\,d\theta \\
      &\approx 6\,\mathrm{SNR}^2\ln\mathrm{SNR}.
\end{split}
\end{equation}
The integral diverges logarithmically at small $\theta$ and is cut off at $\epsilon\sim\gamma$, since pulsars closer to the source than the localization error are fully de-phased. Each component of the angular error has variance $1/F$, so the radial error follows a Rayleigh distribution with median $\sqrt{2\ln2/F}$. This gives the full scaling relation, including the pre-factor:
\begin{equation}
    \gamma_{\rm med} \simeq \sqrt{\frac{\ln 2}{3}}\, \frac{1}{\mathrm{SNR} \sqrt{\ln \mathrm{SNR}}}\ {\rm rad} \approx \frac{27.5^\circ}{\mathrm{SNR}\sqrt{\ln\mathrm{SNR}}}.
    \label{eq:scaling}
\end{equation}

\subsection{Uniform PTA simulation}
\label{sec:uniform_sim}

We verify this scaling relation by simulating uniform, dense pulsar array maps with a single, left circularly polarized point source and random Gaussian noise added in the timing residual space. To approximate an idealized, fully dense array, we place one pulsar at the center of each HEALPix pixel at resolution $N_{\rm side}^{\rm PTA}=20$. This gives 4800 pulsars distributed uniformly over the sky. The matched-filter map is pixelized at a finer resolution of $N_{\rm side}^{A}=100$, which corresponds to a pixel size of $\approx 0.6^\circ$. This is small enough that pixelization does not dominate the angular error at the SNRs of interest.

We inject a single GW point source on the equator. Both the pulsar positions and the map pixels lie on a discrete grid. To keep the result from depending on where the source falls relative to that grid, we randomize the source's azimuthal coordinate $\phi$ independently for each noise realization. The resulting precision is therefore an average over source positions, representative of the sky-averaged performance of a uniform array. For each realization, we add random complex Gaussian noise to the timing residuals and construct the noisy matched-filter map $L$. We estimate the source position as the pixel with the maximum value of $|L|$. The angular error for that realization is the angular separation between this pixel and the true source position. We then compute the median angular error over $N_{\rm real}=500$ noise realizations. To map out the dependence on SNR, we use the same 500 noise realizations at every SNR. We vary the SNR only by rescaling the amplitude of the noiseless map $\mathbf{L}^{(s)}$ according to Eq.~\eqref{eq:snr}, with the noise average taken over the $N_{\rm real}=500$ realizations at the pixel closest to the GW position. Because the noise realizations are held fixed, the variation between SNR values is smooth and reflects the change in signal amplitude rather than differences in noise draws.

\begin{figure}[tbp]
    \centering
    \includegraphics[width=\linewidth]{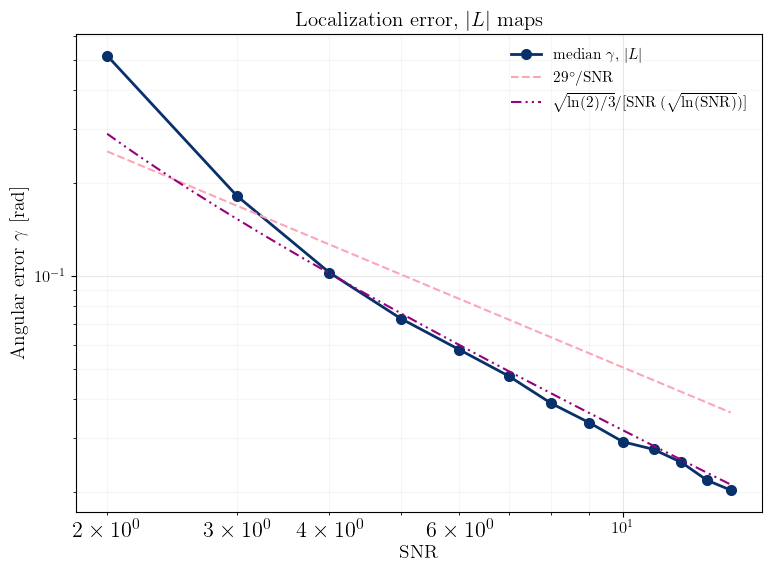}
    \caption{Median angular error as a function of SNR for a uniform PTA of 4800 pulsars, representing the intrinsic astrometric precision for an incoherent PTA. Each point is computed from 500 noise realizations; see Sec.~\ref{sec:uniform_sim} for details. Lines show the Gaussian-PSF scaling $\gamma_{1/2}/\mathrm{SNR}$ with $\gamma_{1/2}=28.9^\circ$ \cite{Jow2026}, and the derived relation $\sqrt{\ln2/3}\,/[\mathrm{SNR}\sqrt{\ln\mathrm{SNR}}]$ of Eq.~\eqref{eq:scaling}. The simulation follows Eq.~\eqref{eq:scaling} at high SNR, giving $\approx 3.3^\circ$ at $\mathrm{SNR}=6$, a factor of about $1.4$ better precision than predicted from a Gaussian PSF.}
    \label{fig:SNR}
\end{figure}

Figure~\ref{fig:SNR} shows the median angular error as a function of SNR. We compare it with two scaling relations: $\gamma_{1/2}/\mathrm{SNR}$, the expected scaling for a Gaussian PSF, normalized by the PSF width $\gamma_{1/2}=28.9^\circ$ found in Ref.~\cite{Jow2026}; and Eq.~\eqref{eq:scaling}, derived in Sec.~\ref{sec:scaling}. At low SNR, the maximum pixel is often a noise fluctuation that can lie anywhere on the sky, generating a long tail for the distribution of angular error values. For the full sky, this tail is bounded by $\pi$. On a flat sky, this tail would be unbounded and angular error values could be arbitrarily large at noise-dominated SNR levels. This feature of the distribution of angular error values motivates our choice to use the median angular error over the root-mean-square (RMS). The median is a more robust measure of the typical error in this case, as it is less sensitive to this geometry-bounded tail. Because the angular error at low SNR is set by the sky geometry rather than by the PSF, we do not expect any particular scaling relation to hold in this regime. At high SNR, the maximum pixel lies within the main lobe of the PSF, and the simulated precision follows the $\sqrt{\ln2/3}\,/[\mathrm{SNR}\sqrt{\ln\mathrm{SNR}}]$ scaling of Eq.~\eqref{eq:scaling}. At $\mathrm{SNR}=6$, the astrometric precision is approximately $3.3^\circ$, a factor of approximately $1.4$ smaller angular error than the Gaussian-PSF prediction.

\section{IPTA pulsar configuration}
\label{sec:IPTA}

In this section, we compare the localization performance of the IPTA DR2 \cite{Perera2019} pulsar distribution with the sky-averaged maximum astrometric precision from the intrinsic PTA response discussed in the previous section. We consider a fixed map SNR of 6, corresponding to a detection significance of $5.2\sigma$ with a maximum-pixel estimator after accounting for the look-elsewhere penalty [see Eq.~(30) of Ref.~\cite{Jow2026}]. We take this as the approximate relevant noise level for current PTA analyses, because applying the stochastic, isotropic Hellings--Downs estimator to the same single-source noisy map would yield a detection significance of $3\sigma$, similar to the current sensitivity of PTAs to the isotropic background. For each injected GW, the SNR is fixed by rescaling the noise-free map so that its maximum pixel value satisfies Eq.~\eqref{eq:snr}. We note again that the noise is averaged over different noise realizations of the same map pixel, not over the sky, because the peak of the point-spread function takes different values across the sky for a non-uniform pulsar distribution.

\begin{figure}[tbp]
    \centering
    \includegraphics[width=\linewidth]{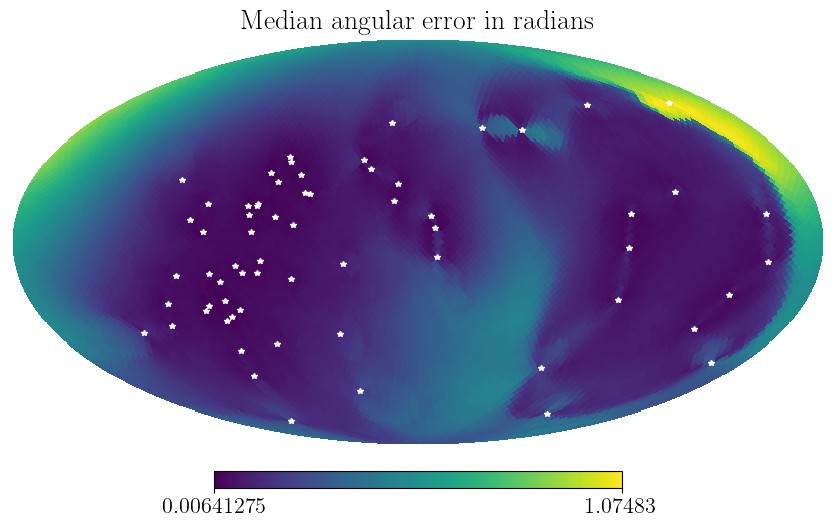}
    \caption{Median angular error for individual GW sources injected at the
    centers of HEALPix pixels with $N_{\rm side}^{\rm GW}=30$. For each injected
    source, we generate $|L|$ maps with $N_{\rm side}^{A}=40$ for 200 noise
    realizations at a fixed SNR of 6, set using the RMS of the noise over
    realizations at the pixel closest to the injected GW position. The angular
    error is again the angular separation between the true GW position and the
    location of the maximum pixel in the noisy map. Pulsar positions match the
    latest completed IPTA data release, DR2~\cite{Perera2019}.}
    \label{fig:IPTA}
\end{figure}

We calculate the astrometric precision by generating $N_{\rm real}=200$ single-source matched-filter maps with the IPTA pulsar distribution, again with random Gaussian noise added to the PTA residuals. We inject point sources at the centers of the $12 (N_{\rm side}^{\rm GW})^2 = 10800$ HEALPix pixels with $N_{\rm side}^{\rm GW} = 30$, and generate matched-filter maps for each injected GW. Figure~\ref{fig:IPTA} shows, for each sky pixel, the median angular error of the maximum pixel of the $|L|$ map over the $N_{\rm real}$ realizations, for a source injected at the center of that pixel. These maps use a pixelization of $N_{\rm side}^{A}=40$ with pixel size $\approx 1.5^\circ$. The discontinuities near the ring of pulsars on the right of Fig.~\ref{fig:IPTA} are due to the high-density region of pulsars on the left of the map, which causes a bimodal distribution of angular error in the right hemisphere by dragging the maximum pixel closer to the region of highest pulsar density in some realizations.

\begin{figure}[tbp]
    \centering
    \includegraphics[width=\linewidth]{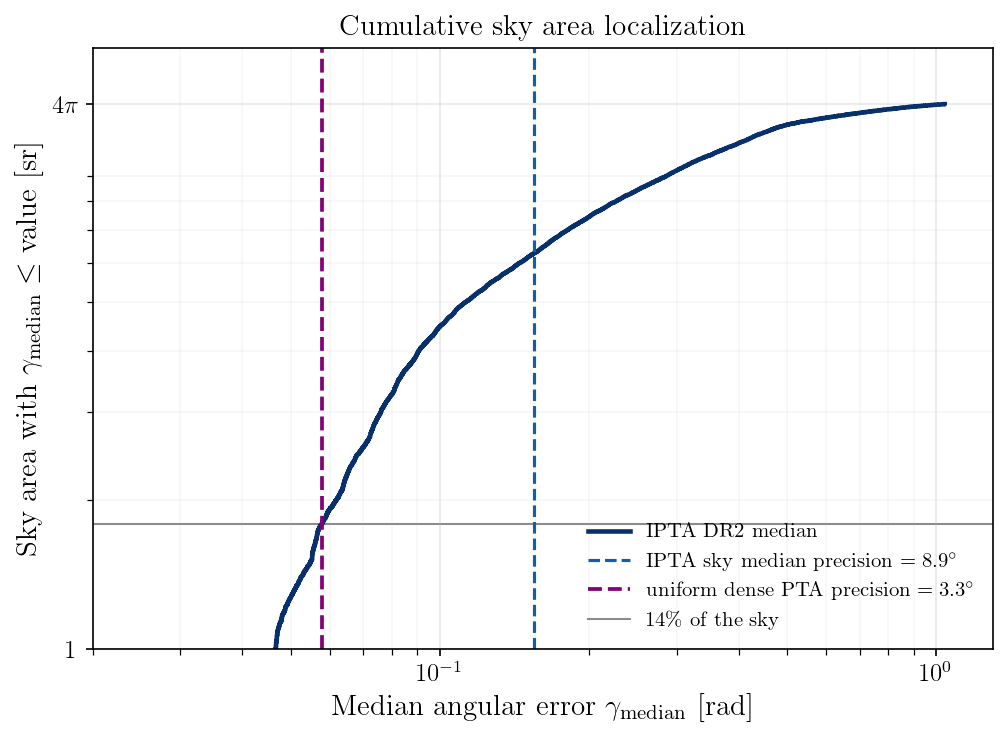}
    \caption{Cumulative sky area over which the IPTA array achieves a
    median angular error of $\gamma_{\rm med}$ or better at $\mathrm{SNR}=6$.
    Vertical dashed lines mark the IPTA sky median ($8.9^\circ$) and the
    intrinsic precision of the uniform, dense PTA ($3.3^\circ$); the horizontal
    line marks the $14\%$ of the sky in which the IPTA array outperforms the
    uniform array.}
    \label{fig:CDF}
\end{figure}

Next, we take the sky map in Fig.~\ref{fig:IPTA} and calculate the cumulative distribution function (CDF) of astrometric precision across the sky, shown in Fig.~\ref{fig:CDF}. We find that the median precision across the sky is approximately $8.9^{\circ}$, compared to the intrinsic precision of $3.3^{\circ}$. We also find that for point sources in about $14\%$ of the sky, the astrometric precision is better than the $3.3^{\circ}$ intrinsic limit due to the relatively high pulsar density in those regions. We note that this precision in $14\%$ of the sky is to the detriment of the rest of the sky where localization is far worse than from the intrinsic response. We discuss this effect in more detail in the next section.

\section{Discussion}
\label{sec:discussion}

We have used a dense, uniform pulsar distribution to approximate the intrinsic, incoherent PTA astrometric precision. This pulsar configuration yields the same precision everywhere on the sky and is the best-case configuration when there is no a priori information on source angular position. What the analysis in Sec.~\ref{sec:uniform} reveals is that the singular core of the convolution kernel shown in Fig.~5 of Ref.~\cite{Jow2026} yields an extra factor of $\sqrt{\ln\mathrm{SNR}}$ improvement on source localization compared to the naive guess of a Gaussian convolution kernel.

Existing literature on nanohertz source localization, Ref.~\cite{Taylor2026}, shows
that the angular localization error of a source scales linearly with the source--pulsar angular separation when considering the PTA response from a ring of pulsars centered at the GW position with varying radius. Pulsars concentrated near the source therefore localize it better than a
uniform array at the same map SNR. This gain is difficult to leverage in practice with current observational information, mainly because source positions are unknown. In a blind
search, the position of the brightest resolvable SMBHB is not known a priori, and
a configuration that is favorable for one direction is unfavorable for most of
the sky: concentrating pulsars in one region necessarily removes them from
others, so the sky-averaged precision of any non-uniform array is worse than
that of a uniform array at the same SNR, as we find for the IPTA configuration
in Sec.~\ref{sec:IPTA}. We are also limited in pulsar configuration by the Galactic population of millisecond pulsars and by which of
them can be timed with sufficient precision. Here, we have studied the intrinsic PTA response, approximated by a dense, uniform array, because it gives the best sky-averaged precision. It also shows how the singular core of the convolution kernel improves localization, as seen in the SNR scaling of Sec.~\ref{sec:uniform}. With no prior knowledge of the source position, this is the best achievable precision. However, the configuration-dependent gains outlined in Ref.~\cite{Taylor2026} could become possible to leverage once a candidate source has been identified.

While the average astrometric precision of an incoherent PTA can be improved by increasing the overall SNR, in previous work, we showed that including precise pulsar distances yields approximately an order of magnitude better source localization per distance, and could reach diffraction-limited localization of approximately $1$ arcmin with 10 distances \cite{Tsai2026}. Thus, although incoherent PTAs follow a sharper scaling relation for source localization than a Gaussian PSF, the orders-of-magnitude better astrometric precision that may be required for host-galaxy identification relies on subparsec-precision pulsar distance measurements.

\section{Conclusion}

In this paper, we study the astrometric precision of an incoherent PTA, in which pulsar distances are too poorly constrained for the pulsar term to contribute significantly to localization. Using a heuristic and first-principles argument, verified by matched-filter map-making simulations of a dense, uniform array, we find that the median angular error scales with map SNR as $\sqrt{\ln2/3}\,/[\mathrm{SNR}\sqrt{\ln\mathrm{SNR}}]$ at high SNR, which is sharper than the scaling expected for a Gaussian beam. This relation represents the best sky-averaged source localization achievable by any incoherent PTA with no a priori knowledge of source locations, approximately $3.3^{\circ}$ at a map SNR of 6. The IPTA DR2 pulsar configuration performs worse at the same SNR, with a median astrometric precision across the sky of about $8.9^\circ$. However, sources in the pulsar-dense $14\%$ of the sky can be localized better than in the uniform case. Reaching the much finer precision potentially needed for host-galaxy identification will require a coherent analysis that uses precise pulsar distances \cite{Tsai2026}.

\section*{Acknowledgments}

U.P. is supported by the Natural Sciences and Engineering Research Council of Canada (NSERC) [funding reference number RGPIN-2019-06770, ALLRP 586559-23, RGPIN-2025-06396], Ontario Research Fund (ORF-RE Fund), and AMD AI Quantum Astro, the Academia Sinica [Grand Challenge Seed Program AS-GCS-114-M02 and Investigator Award AS-IV-115-M04], and the National Science and Technology Council (NSTC) of Taiwan [grant no. 115-2123-M-001-009]. We used Claude (Anthropic) to assist with drafting and editing portions of the manuscript text, checking internal consistency, and developing and checking the analytic derivation of the scaling relation in Sec.~\ref{sec:scaling}, which we verified independently against our simulations. The authors reviewed all AI-assisted content and take full responsibility for the final manuscript.

\bibliography{references}

\end{document}